\documentclass[useAMS,usenatbib]{mn2e}
\usepackage{natbib}
\usepackage{amsmath}
\usepackage{subfig}
\usepackage{graphicx}
\usepackage{url}
\usepackage[usenames,dvipsnames]{xcolor}
\usepackage{float}

\title[Monthly Notices: \LaTeXe\ guide for authors]
  {Application of chaos indicators in the study of dynamics of S-type extrasolar planets in stellar binaries}
\author[Satyal et al.]
  {S.~Satyal,$^1$
  B.~Quarles,$^1$\thanks{e-mail: bquarles@uta.edu} T.C.~Hinse$^{2,3}$ 
  \\
  $^1$Department of Physics, University of Texas at Arlington, Arlington,
      Texas, 76019, USA.\\
  $^2$ Korea Astronomy and Space Science Institute, 304-358 Daejeon,
Republic of Korea.\\
  $^3$ Armagh Observatory, College Hill, BT61 9DG, Armagh, UK.}
\date{Released 2013 March 22}

\pagerange{\pageref{firstpage}--\pageref{lastpage}} \pubyear{2002}

\def\LaTeX{L\kern-.36em\raise.3ex\hbox{a}\kern-.15em
    T\kern-.1667em\lower.7ex\hbox{E}\kern-.125emX}

\begin{document}

\label{firstpage}

\maketitle

\begin{abstract}
 The orbits of two individual planets in two known binary star systems, $\gamma$ Cephei and HD 196885 are numerically integrated using various numerical techniques to assess the chaotic or quasi-periodic nature of the dynamical system considered. The Hill stability (HS) function which measures the orbital perturbation of a planet around the primary star due to the secondary star is calculated for each system. The maximum Lyapunov exponent (MLE) time series are generated to measure the divergence/convergence rate of stable manifolds, which are used to differentiate between chaotic and non-chaotic orbits. Then, we calculate dynamical Mean Exponential Growth factor of Nearby Orbits (MEGNO) maps from solving the variational equations along with the equations of motion. These maps allow us to accurately differentiate between stable and unstable dynamical systems. The results obtained from the analysis of HS, MLE, and MEGNO maps are analysed for their dynamical variations and resemblance. The HS test for the planets shows stability and quasi-periodicity for at least ten million years. The MLE and the MEGNO maps have also indicated the local quasi-periodicity and global stability in a relatively short integration period. The orbital stability of the systems is tested using each indicator for various values of planet inclinations ($i_{pl}$$\le$25$^\circ$) and binary eccentricities. The reliability of HS criterion is also discussed based on its stability results compared with the MLE and MEGNO maps.
\end{abstract}


\section{Introduction}

The discovery of extra solar planets has been growing substantially since the first planet, 51 Pegasi b, was detected almost two decades ago \citep{may95}. Since then 872 extra solar planets have been confirmed as of April 25, 2013\footnote{\url{www.exoplanet.eu}}. Near half of solar type stars \citep{duq91,rag06} and a third of all stars in the Galaxy \citep{rag10} are in a binary or multi-star system with 40 planets confirmed in such systems \citep{des07}. The confirmation of the existence of planets in binaries has raised a new astrophysical challenge which includes the study of long term orbital stability of such planets. The ultimate destiny of exoplanet research, including observations from the Kepler mission\footnote{\url{www.nasa.gov/mission_pages/kepler/overview/index.html}}, is to detect a planet on a stable orbit within the habitable zone of the host star \citep{bor97,koc07,bor08,bor10}. The degree of stability is largely governed by the planet's semimajor axis, eccentricity and orbital inclination. Orbital long-term stability is believed to be a necessary condition for life to develop. The study of orbital stability furthers one self-evident aim of mankind which is to find an answer to the century old question, ``Are we alone in the Milky Way Galaxy?''.

By using different stability criteria the question of stability has been addressed by many others in the past. While studying the Trojan type orbits around Neptune, \cite{zho09} showed that the inclination of orbits can be as high as 60$^\circ$ while maintaining orbital stability. Several authors \citep{sze80,sze81,sze08} calculated orbital stability of planets by using several techniques that include the integrals of motion, zero velocity surfaces (ZVS), and a Hill stable region that is mapped by a parameter space of orbital radius and mass ratio, $\mu$, for a coplanar circular restricted three body problem (CRTBP). \cite{qua11} has also used the maximum Lyapunov exponent (MLE) to determine the orbital stability or instability for the CRTBP case. The stability limits were defined based on the values of MLE that are dependent on the mass ratio $\mu$ of the binaries and the initial distance ratio $\rho_0$ of the planet. Other chaos indicator techniques such as Mean Exponential Growth factor of Nearby Orbits (MEGNO) maps have also been used to study the dynamical stability of irregular satellites \citep{hin10} and extrasolar planet dynamics \citep{goz01a,goz01b}. The MEGNO criterion is known to be efficient in distinguishing between chaotic and quasi-periodic initial conditions within a dynamical phase space.

Knowing the orbital stability of planets is a crucial step for further studies of planetary systems. In order to probe a planet for its habitability, there exists a primary requirement that the system be orbitally stable. In this paper, we have used three stability indicators, HS, MLE, and MEGNO maps, in the study of the orbital dynamics of planets in the selected stellar binaries. We have also compared all the results from three different methods in order to determine if HS can be considered a reliable, efficient tool in the stability analysis of exoplanets.

This paper is outlined as follows. In Section 2 we discuss the basic theory of the analysis tools. In Section 3 we present our numerical methods and the results of the stability analysis from different chaos indicators followed by discussion. Finally, we will conclude in Section 4 with a brief overview of our results.

\section{Theory}
\subsection{Basic Definitions and Equations}

For the motion of a planet of mass $m_1$ around a star of mass $m_2$ in an orbital plane, the initial velocity was calculated using the time derivative of the position matrix given by \cite{mur99}. While calculating the initial conditions we used the values of the parameters ($a$, $e$, $i$ $\Omega$, $\omega$ and $f$) whenever they have been observationally determined.

Our particular interest is on the stellar binaries that are less than or equal to 25 AU apart. For the stars with greater than 25 AU separations, the effects of the secondary star on the planet would not be significant, especially while considering the intent of our present study.

The list of planets in the binaries and their orbital parameters are given in Tables \ref{tab:gCephei} and \ref{tab:HD196885}. The initial setup in our simulations is in a barycentric coordinate system with the appropriate placement of Star B (left) and Star A (right) relative to the barycentre.  The positive x-axis is taken to be the reference.  The true anomaly of both the binary and planet are assumed to be equal to zero.

For two primaries in elliptic orbits moving about their barycentre the dynamical system of a third smaller mass can be written as the first order differential equations. The equations of motion are given below \citep{sze67, sze08} \\
\begin{align}
x' &= u& u' &= 2v + {1\over(1+e\cos{f})}\left[x-{\alpha (x+\mu) \over r_1^3} - {\mu(x-1+\mu) \over r_2^3}\right], \nonumber \\
y' &= v& v' &= -2u +{y\over(1+e\cos{f})}\left[1-{\alpha  \over r_1^3} - {\mu \over r_2^3}\right], \nonumber \\
z' &= w& w' &= -z + {z\over(1+e\cos{f})}\left[1-{\alpha  \over r_1^3} - {\mu \over r_2^3}\right],
\label{eqn:diffeqn}  
\end{align}
where
\begin{align}
\mu &= {m_2\over(m_1+m_2)},\nonumber \\
\alpha &= 1-\mu,\nonumber \\
r_1^2 &= {(x-\mu)}^2 + y^2 + z^2,\nonumber \\
r_2^2 &= {(x+\alpha)}^2 + y^2 + z^2.
\end{align}
 
The Jacobi constant ($C_0$) for a initial state (\emph{$x_o$, $y_o$, $z_o$, $f_o$}) is given by \\
\begin{align}
C_0 = {x_0^2 + y_0^2 + {2{(1-\mu)}\over {r_1}} + {2{\mu}\over {r_2}} \over 1+e\cos(f_0)} -\dot{x}_0^2 - \dot{y}_0^2 - \dot{z}_0^2.
\end{align}

In Eqn \ref{eqn:diffeqn}, the variables represent the velocity of a test particle (planet) in Cartesian coordinates (\emph{x, y, z}). The distances $r_1$ and $r_2$ are defined in terms of mass ratio, normalised coordinates, and the position of the stars within a pulsating-rotating coordinate system\footnote{Although the masses were integrated using a N-Body representation, we have included the ERTBP equations of motion to illustrate the necessary transformations to arrive at the pulsating-rotating coordinate system (see \cite{sze67} for full details) used in our analysis.}. 

\subsection{Lyapunov Exponents}
For stable planetary orbits, the two nearby trajectories in phase space will converge and for unstable orbits, the trajectories diverge exponentially. The rate of divergence is measured by using the method of Lyapunov exponents \citep{lya07}. \cite{wol85} developed a numerical method of computing the Lyapunov exponents in FORTRAN following the earlier works by \cite{ben80}.
 
Lyapunov exponents are commonly used because they give the measure of an attractor of a dynamical system  as it converges or diverges in phase space. The positive Lyapunov exponents measure the rate of divergence of neighbouring orbits, whereas negative exponents measure the convergence rates between stable manifolds \citep{tso92,ott93}. The sum of all Lyapunov exponents is less than zero for dissipative systems \citep{mus09} and zero for non-dissipative (Hamiltonian) systems \citep{hil94}. Lyapunov exponents for the circular restricted three body problems (CRTBP) have been calculated previously \citep{gon81,mur01} and similar methods have been used in the elliptic restricted three body problem (ERTBP). In this work, we have calculated the Lyapunov exponents for the N-body problem where $N=3$. In order to calculate the Lyapunov exponents a dynamical system with \emph{n} degrees of freedom is represented in a \emph{2n} phase space. Then the state vectors (\emph{2n}) containing 6 elements are used to calculate the Lyapunov exponents. The details on the calculation of Jacobian J from the equations of motion can be found in \cite{qua11}.

For a Hamiltonian system (see above) to be stable, the sum of all the Lyapunov exponents should be zero. In order to numerically meet such a criterion, a simulation of the system would require an impractically long period of time. Within the limits of simulation, the sum of all six exponents must remain numerically close to zero ($\sim\;10^{-10}$). We have used the largest positive Lyapunov exponent to determine the magnitude of the chaos while ignoring the lesser positive and the negative exponents as they do not provide significant additional information about the evolving system. The positive maximum Lyapunov exponent is known to indicate a chaotic behaviour in both dissipative \citep{hil94} and non-dissipative \citep{ozo90} systems. Chaos can be proven up to the integration time if a given chaos indicator has converged to an unstable manifold.

\begin{table}
\centering
\begin{tabular}{|l|c|c|c|}
\hline
\textbf{$\gamma$ Cephei} & \textbf{A} & \textbf{B} & \textbf{Ab} \\
\hline
Mass               & 1.4 M$_\odot$ & 0.362 M$_\odot$ & 1.6 M$_J$ \\
Semimajor Axis (a) & 19.02 AU	   &                 & 1.94 AU   \\
Eccentricity (e)   & 0.4085        &                 & 0.115     \\
Argument of Periapsis ($\omega_p$) & 0$^\circ$ & 180$^\circ$       & 94$^\circ$ \\
\hline
\end{tabular}
\caption{Orbital parameters of $\gamma$ Cephei (Neuhauser et. al. 2007)}
\label{tab:gCephei}

\begin{tabular}{|l|c|c|c|}
\hline
\textbf{HD 196885} & \textbf{A} & \textbf{B} & \textbf{Ab} \\
\hline
Mass               & 1.33 M$_\odot$ & 0.45 M$_\odot$ & 2.98 M$_J$ \\
Semimajor Axis (a) & 21 AU	   &                 & 2.6 AU   \\
Eccentricity (e)   & 0.42        &                 & 0.48     \\
Argument of Periapsis ($\omega_p$) & 0$^\circ$ & 180$^\circ$       & 93.2$^\circ$ \\
\hline
\end{tabular}
\caption{Orbital parameters of HD 196885, \citep{cha07}}
\label{tab:HD196885}
\end{table}

\subsection{Hill Stability}
\cite{Hil78b,Hil78a} developed the equations of motion for a particle around the primary mass in presence of a nearby secondary mass. The purpose of the Hill equations was to calculate the orbital perturbation of the particle due to the secondary mass. Later the idea was further developed and used in the study of orbital stability of planets \citep{sze67,wal81,mar82}.

The significant radial gravitational influence of the secondary mass reaches as far as the Lagrange points, L1 and L2, forming the Hill sphere \citep{Hil78b,Hil78a}. The contour lines within the sphere are the zero velocity curves. After measuring a particle's position and velocity, a constant of motion relation can be implemented \citep{sze67,mur99}  2$U$ - $v^2$ = $C_J$, where $v$ is the velocity, $U$ is the generalised potential, and $C_J$ is the constant of integration called the Jacobi constant. When the velocity of the particle is zero, 2$U$ = $C_J$, a contour represents a zero velocity surface (ZVS) and the motion of a particle within such a surface is considered Hill stable.

The measure of Hill stability for the ERTBP, $S(f)$, is given by a parameter dependent potential, $\Omega(\textbf{x},f)$, where $f$ is the true anomaly and $C_{cr}$ is the value of the Jacobi constant at the Hill radius or the Lagrange point L1 \citep{sze08}

\begin{align}
\Omega(x,y,z,f) &= {1 \over 2}\left[x^2 + y^2 - ez^2\cos(f)\right] + {1 - \mu \over r_1^2} + {\mu \over r_2^2} \nonumber \\
&+ {1 \over 2} \mu(1-\mu), \nonumber \\
S(f) &= {1 \over C_{cr}}\left[2\Omega(\textbf{x},f)-\left[1+e\cos(f)\right]v^2\right]-1.
\end{align}

Using the orbital parameters obtained from the numerical integration the potential, $\Omega(\textbf{x},f)$, is calculated to obtain the Hill stability function $S(f)$.  Although the Hill stability function depends on the true anomaly, it can also be represented as a time series.  We have implemented this representation in our results concerning the Hill stability function. When the measure of $S(f)$ of a planet is positive then we have the indication of quasi-periodic orbits and its motion is Hill stable. But when the measure of 
$S(f)$ is negative then the planet enters the instability region.

\begin{figure*}
\centering
\subfloat{\includegraphics[width=.23\linewidth]{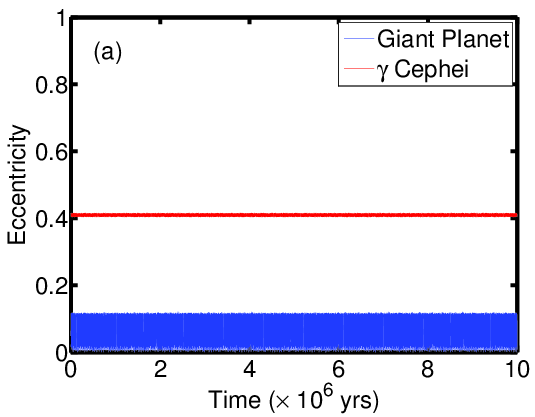}}
\subfloat{\includegraphics[width=.23\linewidth]{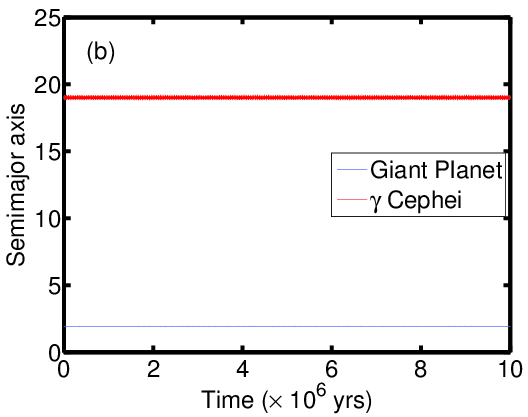}}
\subfloat{\includegraphics[width=.23\linewidth]{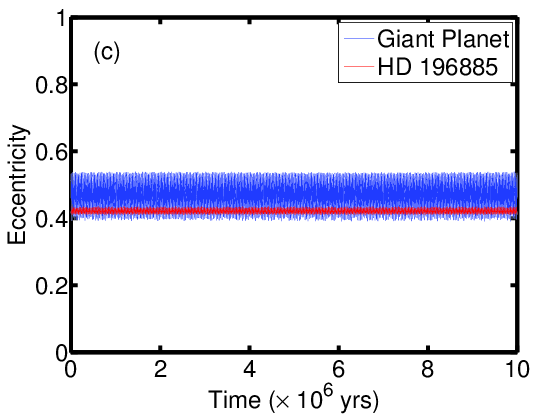}}
\subfloat{\includegraphics[width=.23\linewidth]{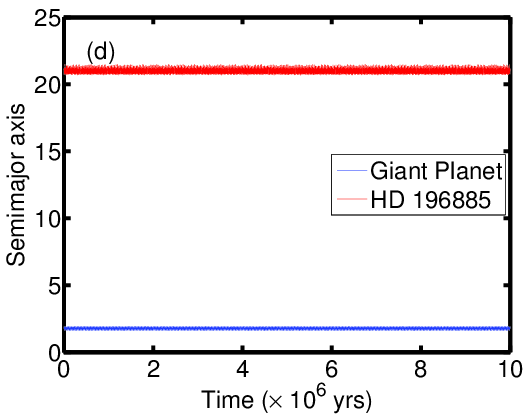}}
\caption{Variation of orbital elements for the giant planets in $\gamma$ Cephei and HD 196885 ($i_{pl}$ = 0.0) simulated for $1 \times 10^7$ years.}
\label{fig:gCephei_ecc}
\end{figure*}

\subsection{The MEGNO Chaos Indicator}
The MEGNO criterion was first introduced by \cite{cin99,cin00,cin03} and found wide-spread applications in dynamical astronomy \citep{goz01b,goz08,goz09,hin08,hin10,fro11,com12,kos12}. The MEGNO (usually denoted as $\langle Y \rangle$) formalism has the following mathematical properties. In general, MEGNO has the parameterisation $\langle Y \rangle = \alpha \times t + \beta$ (see references above). For a quasi-periodic initial condition, we have $\alpha \simeq 0.0$ and $\beta \simeq 2.0$ (or $\langle Y \rangle \rightarrow 2.0$) for $t \rightarrow \infty$ asymptotically. If the orbit is chaotic, then $\langle Y\rangle \rightarrow \lambda t/2$ for $t \rightarrow \infty$. Here $\lambda$ is the maximum Lyapunov exponent (MLE) of the orbit. In practice, when generating our MEGNO maps, we terminate a given numerical integration of a chaotic orbit when $\langle Y \rangle > 12.0$. Quasi-periodic orbits have $|\langle Y\rangle - 2.0| \le 0.001$.

We used the MECHANIC\footnote{http://www.git.astri.umk.pl/projects/mechanic} software \citep{slo12} optimised to N-body code to calculate the orbits of the given masses and the MEGNO maps on a multi-CPU computing environment. Typically we allocated 60 CPUs for the calculation of one map considering a typical grid of $(500\times300)$ initial conditions in $(a,e)$ space. The numerical integration of the equations of motion and the associated variational
equations \citep{mik99} are based on the ODEX integration software \citep{hai93} which implements a Gragg-Bulirsch-Stoer algorithm. The MEGNO indicator is calculated from solving two
additional differential equations as outlined in \cite{goz01b}.

MEGNO and the maximum Lyapunov exponent (MLE) have a close relation where both indicators provide the magnitude of the exponential divergence of orbits. \cite{fro97} introduced the fast Lyapunov indicator (FLI), which exhibits the least dependency on initial conditions. \cite{mes11} showed that MEGNO and FLI are related to each other. FLI is used to detect weak chaos and is considered a faster means to determine the same characteristics as MLE. Recently, \cite{maf11} compared various chaos indicators including FLI and MEGNO. The MEGNO technique and FLI are considered to be in the same class of chaos detection tools \citep{mor02}, and we have chosen the MEGNO technique to compare against the Hill stability criterion. More on mathematical properties of MEGNO and its relationship with the Lyapunov exponents can be found in \cite{hin10}.

\section{Results and Discussion}

\subsection{Numerical Simulation}
To establish the Hill stability (HS) criterion and calculate the maximum Lyapunov exponents (MLE), we numerically simulated each of the planets in the stellar binaries using a Yoshida sixth order symplectic and a Gragg-Bulirsch-Stoer integration scheme \citep{yos90, gra96,hai93}. A stepping of $\epsilon$=$10^{-3}$ years/step was used in each case to have a  better measure of the precision of the integration scheme. The error in energy was calculated at each step which falls in the range of $10^{-14}$ to $10^{-10}$ during the total integration period. Numerical simulations were completed for a million years to calculate the MLE and 10 million years to calculate the HS. MEGNO maps are calculated using 100,000 years per initial condition. Chaos, quasi-perodicity, or regular motion can be shown up to the integration time; however, the long term evolution of these systems can only be proven for chaos where the other types of motion are inferred due to the moderate presence or lack of chaos.  

The purpose of simulating for 10 million years was to display the evolution of eccentricity and semimajor axis without applying any indicator tools to provide a consistent comparison in order to analyse the orbital behaviour and to establish the full effectiveness of the indicators. The time series plots for the systems, $\gamma$ Cephei and HD 196885, are relatively constant with minor oscillations for 10 million years (Fig. \ref{fig:gCephei_ecc}). In these cases we found that the eccentricity of the giant planets is oscillating with a constant amplitude. For example, Figs. \ref{fig:gCephei_ecc}a and \ref{fig:gCephei_ecc}c demonstrate oscillations from 0 to 0.1 and 0.4 to 0.5 in values of eccentricity for $\gamma$ Cephei and HD 196885, respectively. The amplitude of the oscillation changed with a different choice of initial conditions.  As a result, specific choices can minimise the oscillation amplitude and can render the simulation for $\gamma$ Cephei to be in closer agreement with previous studies by \cite{hag06}. One such initial condition involved the choice of eccentricity. If \emph{e} = 0 for the planetary orbit initially, we observed that the amplitude of oscillation is minimum and consistent with \cite{hag06} while the amplitude increases with larger initial \emph{e} values.

\subsection{MLE: Indicator Analysis} \label{sec:MLE_An}
\begin{figure*}
\centering
\subfloat{\includegraphics[width=.23\linewidth]{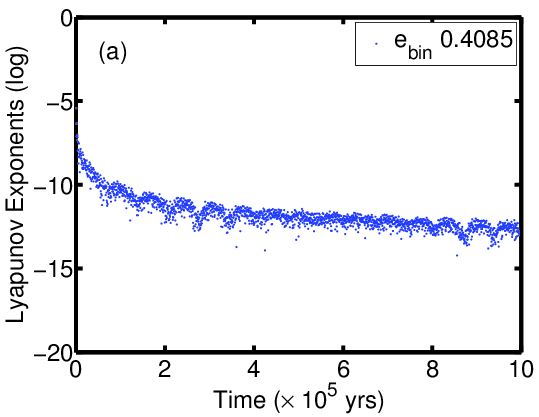}}
\subfloat{\includegraphics[width=.23\linewidth]{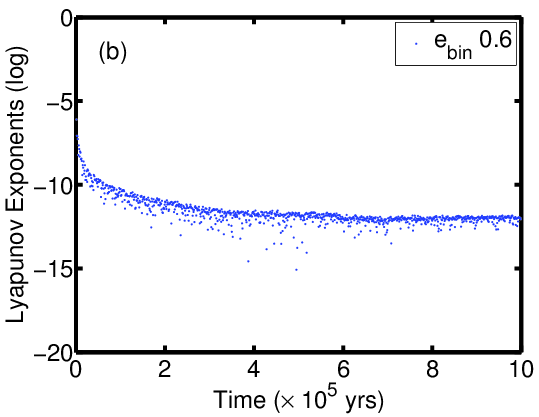}}
\subfloat{\includegraphics[width=.23\linewidth]{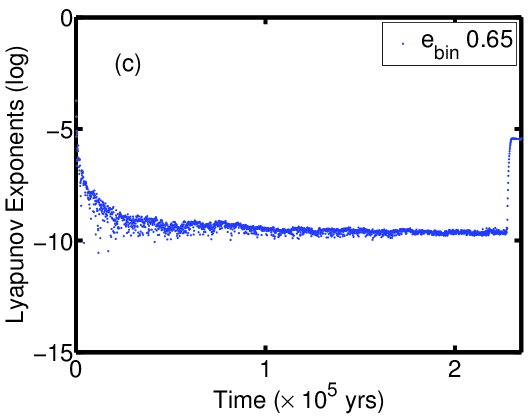}}
\subfloat{\includegraphics[width=.23\linewidth]{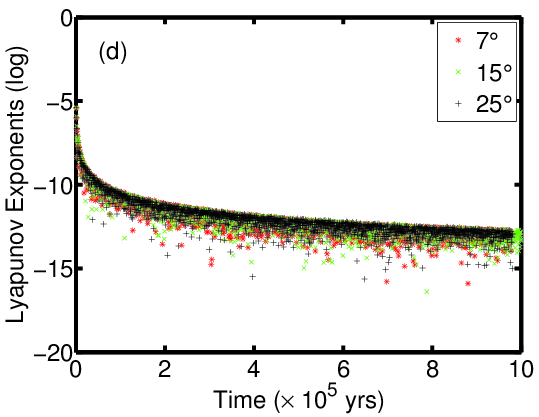}}
\caption{Lyapunov exponent time series for the giant planet in $\gamma$ Cephei with respect to variations in $e_{bin}$ and $i_{pl}$.  The coplanar cases shown (a),(b) and (c) considered initial values of $e_{bin} = 0.4085$, 0.6, and 0.65, respectively. (d) illustrates the time series with the same $e_{bin}$ as in (a) but with different initial values for $i_{pl} = 7^\circ$, 15$^\circ$ and 25$^\circ$ as indicated by the legend. Note: The time axis of (c) has been truncated to the point of instability.}
\label{fig:gCephei_LE}
\end{figure*}

\begin{figure*}
\centering
\subfloat{\includegraphics[width=.23\linewidth]{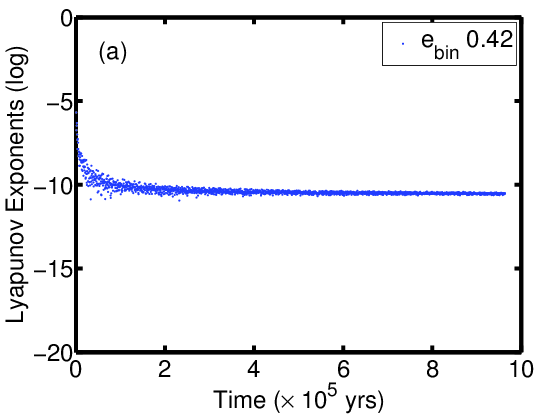}}
\subfloat{\includegraphics[width=.23\linewidth]{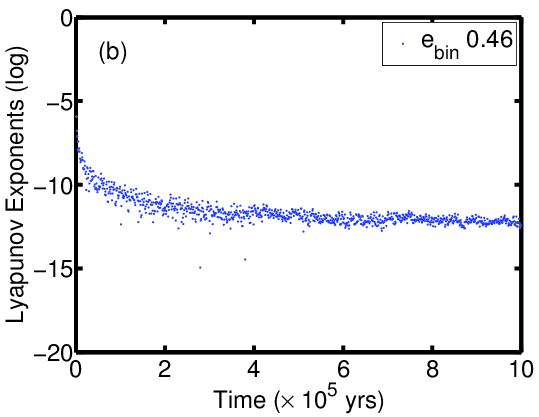}}
\subfloat{\includegraphics[width=.23\linewidth]{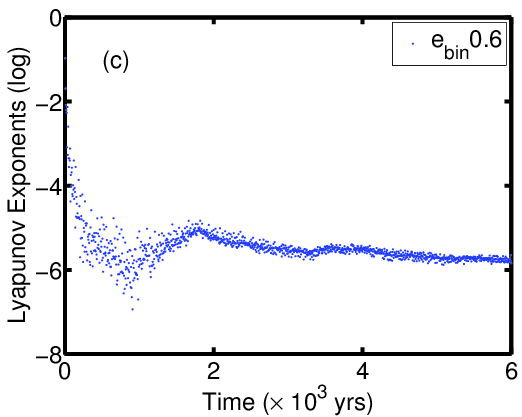}}
\subfloat{\includegraphics[width=.23\linewidth]{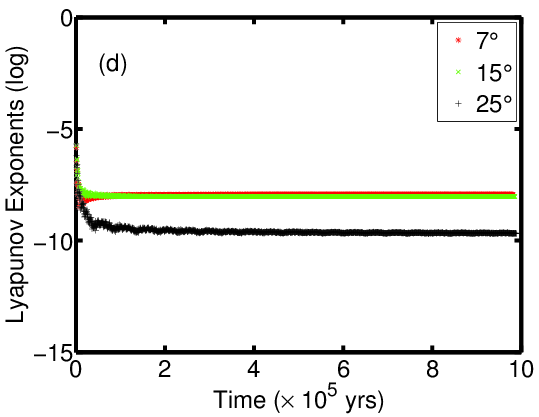}}
\caption{Lyapunov exponent time series for the giant planet in HD 196885 with respect to variations in $e_{bin}$ and $i_{pl}$.  The coplanar cases shown (a), (b) and (c) considered initial values of $e_{bin} = 0.42$, 0.46, and 0.6, respectively. (d) illustrates the time series with the same $e_{bin}$ as in (a) but with different initial values for $i_{pl} = 7^\circ$, 15$^\circ$ and 25$^\circ$ as indicated by the legend. Note: The time axis of (c) has been truncated to illustrate the nature of the instability.}
\label{fig:HD196885_LE}
\end{figure*}

The maximum Lyapunov exponent (MLE) time series for the simulated planets in the stellar binaries are given in Figs. \ref{fig:gCephei_LE} and \ref{fig:HD196885_LE}. The MLE is plotted using a logarithmic scale along the y-axis and a linear scale along the time-axis. We obtained six Lyapunov exponents from our simulation among which three are negative and three are positive. We inspected the first three positive LEs and found the magnitude of the largest value which is used for our purpose of establishing the stability of a system. In Figure \ref{fig:gCephei_LE}, we have taken the maximum Lyapunov exponent as the primary indicator of the orbital stability. For a given initial condition, the MLE must quickly drop below a cut off value similar to \cite{qua11} and decrease at a constant rate where we can determine it as stable or unstable.

The MLE indicates stability for both of the considered systems in the coplanar case, as expected from the observations. This is demonstrated by the values of the MLE which start on the order of 1 and slowly converge down by orders of 10 on a logarithmic scale. The MLE for $\gamma$ Cephei is calculated using three values of initial binary eccentricity: first at the nominal observational value (Fig. \ref{fig:gCephei_LE}a), second within the semi-chaotic region (Fig. \ref{fig:gCephei_LE}b) and third within a region of chaos (Fig. \ref{fig:gCephei_LE}c). The $e_{bin}$ values assumed for the semi-chaotic and chaotic regions were obtained from MEGNO maps (section \ref{sec:MEGNO_An}). While varying the value of $e_{bin}$, the orbital inclination was kept constant at 0$^\circ$. For the first case ($e_{bin}$ = 0.4085), Fig. \ref{fig:gCephei_LE}a shows the MLE slowly decreasing to -13 in a million years. The MLE also shows a similar trend for the second case ($e_{bin}$ = 0.6). Hence, considering this decreasing trend and the nature of Lyapunov exponents (Sect. 2.2), the results reflect the outcome of orbital stability for the planet for our two choices of $e_{bin}$. For the third choice of $e_{bin}$ = 0.65, the MLE time series began displaying instabilities within $2.25 \times 10^5$ years.

The MLE for $\gamma$ Cephei Ab (Fig. \ref{fig:gCephei_LE}d) is also calculated using three additional values of initial orbital inclination ($i_{pl}$ = 7$^\circ$, 15$^\circ$, 25$^\circ$). While varying the initial planetary inclination, $e_{bin}$ was set to the nominal value. Irrespective of our considered initial values in inclination the MLEs of the planet remained unaffected where all cases demonstrate a decreasing trend to -13 in a million years, reflecting the criterion for orbital stability.

Similar to the case of $\gamma$ Cephei, we calculated the MLE for the planet in HD 196885 for three values of binary eccentricity. For the nominal observational value of $e_{bin}$ (Fig. \ref{fig:HD196885_LE}a) and $e_{bin}$ chosen to be at the semi-chaotic region (Fig. \ref{fig:HD196885_LE}b), the MLEs clearly display stable orbits. But, as the $e_{bin}$ was raised to 0.6 the MLE time series displayed an orbital instability within $1 \times 10^3$ years (Fig. \ref{fig:HD196885_LE}c), as also shown by the MEGNO result (see Fig. \ref{fig:HD196885_MEGNO1}c).

The MLE time series for HD 196885 Ab, Fig. \ref{fig:HD196885_LE} is distinct for all considered initial values of the planet's inclination, $i_{pl}$, and the orbit of the planet exists in a stable configuration. However when using our four choices of $i_{pl}$ values, the MLE time series demonstrates that the planetary orbits are more perturbed when its positioned at 7$^\circ$ and 15$^\circ$. The MLE values drop quickly to -7 and levels off indicating chaotic behaviour but does not lead to instability. This may be due to the effects of a near resonance behaviour. However the stability criterion is not violated, thus this system could exhibit long term chaotic motions akin to Pluto in the Solar System. For $i_{pl}$ = 0$^\circ$ and 25$^\circ$ cases, the MLE values appear to be converging slower up to -10 for the first few thousand years than the other considered cases and indicate stable orbits.

\subsection{Establishing the Hill Stability criterion}

\begin{figure*}
\centering
\subfloat{\includegraphics[width=.3\linewidth]{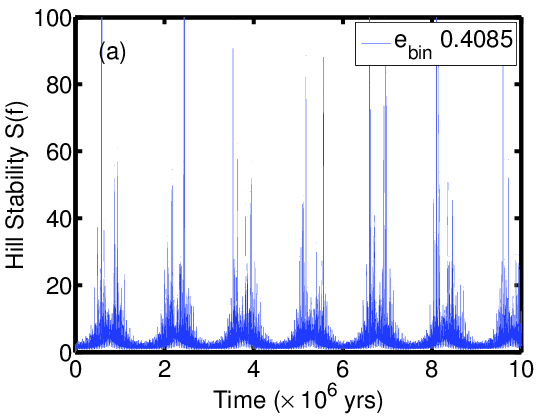}}
\subfloat{\includegraphics[width=.3\linewidth]{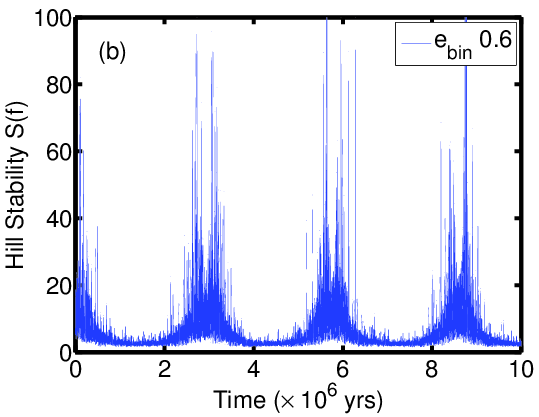}}
\subfloat{\includegraphics[width=.3\linewidth]{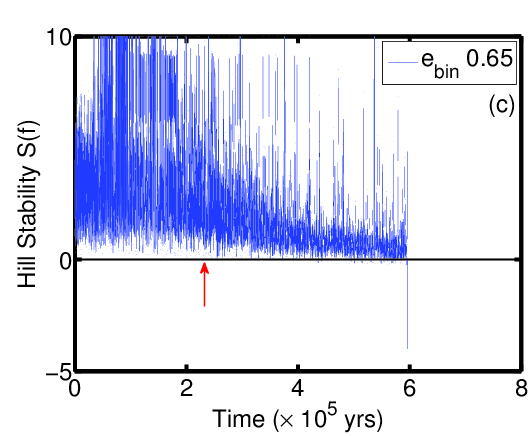}}
\caption{The Hill Stability function for the giant planet in $\gamma$ Cephei for different cases of binary eccentricity (for $i_{pl}$ = 0$^\circ$) simulated for $1 \times 10^7$ years (see Figs. \ref{fig:gCephei_LE}(a), (b) and (c)).  Note: The time axis of (c) has been truncated to illustrate the nature of the instability (shown by the vertical arrow) and the point of ejection.}
\label{fig:gCephei_HS1}
\end{figure*}

\begin{figure*}
\centering
\subfloat{\includegraphics[width=.3\linewidth]{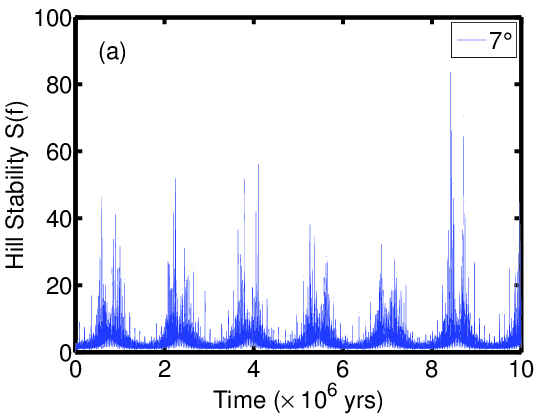}}
\subfloat{\includegraphics[width=.3\linewidth]{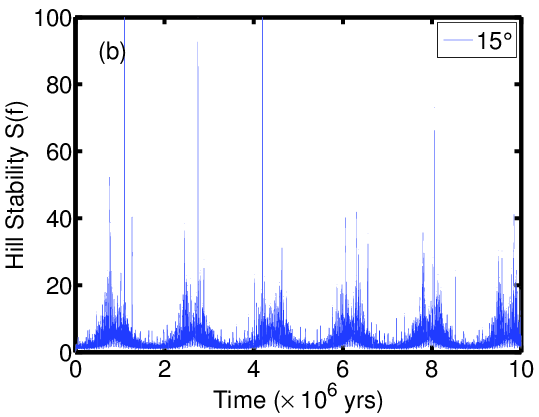}}
\subfloat{\includegraphics[width=.3\linewidth]{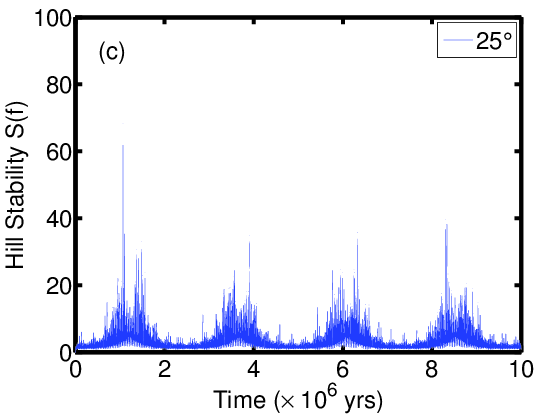}}
\caption{The Hill Stability function for the giant planet in $\gamma$ Cephei for different cases of initial planetary inclination (for $e_{bin}$ = 0.4085) simulated for $1 \times 10^7$ years (see LE time series Fig. \ref{fig:gCephei_LE}(d)).}
\label{fig:gCephei_HS2}
\end{figure*}

\begin{figure*}
\centering
\subfloat{\includegraphics[width=.3\linewidth]{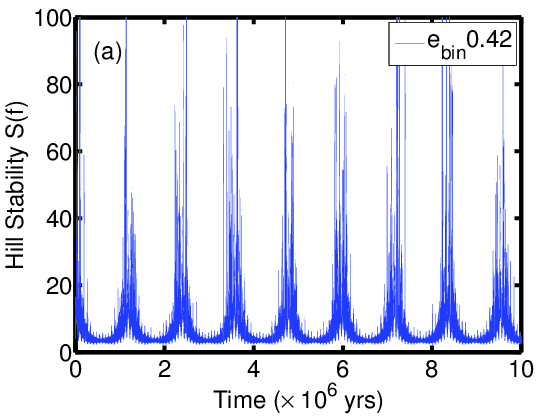}}
\subfloat{\includegraphics[width=.3\linewidth]{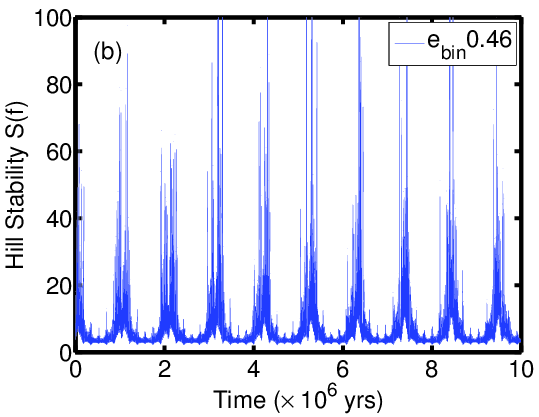}}
\subfloat{\includegraphics[width=.3\linewidth]{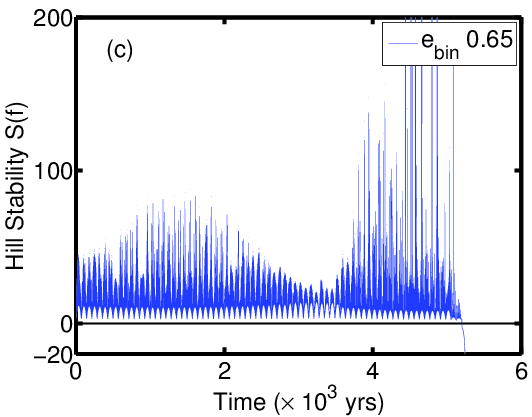}}
\caption{The Hill Stability function for the giant planet in HD 196885 for different cases of binary eccentricity (for $i_{pl}$ = 0$^\circ$) simulated for $1 \times 10^7$ years (see Figs. \ref{fig:gCephei_LE}(a), (b) and (c)).  Note: The time axis of (c) has been truncated to illustrate the nature of the instability and the point of ejection.}
\label{fig:HD196885_HS1}
\end{figure*}

\begin{figure*}
\centering
\subfloat{\includegraphics[width=.3\linewidth]{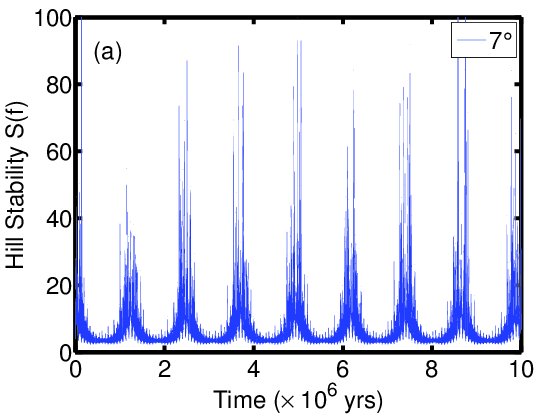}}
\subfloat{\includegraphics[width=.3\linewidth]{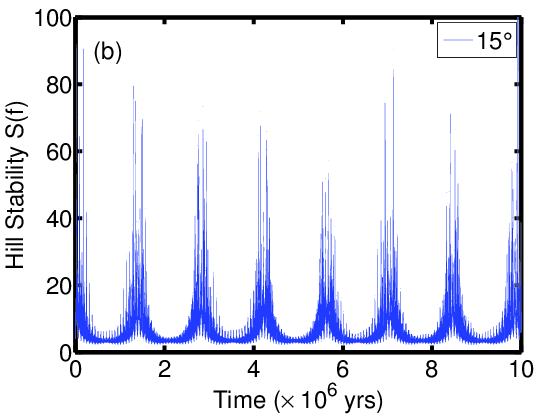}}
\subfloat{\includegraphics[width=.3\linewidth]{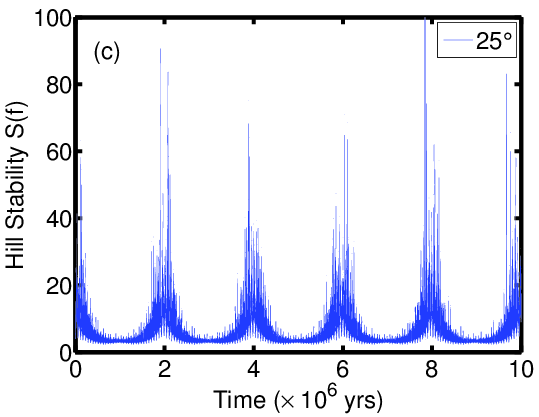}}
\caption{The Hill Stability function for the giant planet in HD 196885 for different cases of initial planetary inclination (for $e_{bin}$ = 0.42) simulated for $1 \times 10^7$ years (see LE times series Fig. \ref{fig:HD196885_LE}(d)).}
\label{fig:HD196885_HS2}
\end{figure*}

The Hill stability time series for the planets in $\gamma$ Cephei and HD 196885 are shown in Figs. \ref{fig:gCephei_HS1}-\ref{fig:HD196885_HS2}. Starting with the coplanar case and the nominal value of binary eccentricity, we have studied the cases when $i_{pl}$ = 7$^\circ$, 15$^\circ$ and 25$^\circ$. We have also calculated the HS functions for two other choices of $e_{bin}$. The choices of our $e_{bin}$ and $i_{pl}$ values are similar to the ones that we made to calculate the MLE time series and MEGNO maps for both of the binary systems.

The measure of Hill stability for $\gamma$ Cephei Ab stays positive for $e_{bin}$ = 0.4085 and $e_{bin}$ = 0.6 throughout the integration period (10 million years) (Figs. \ref{fig:gCephei_HS1}a,b). The oscillations on average are small and positive which reflect the condition for HS criterion established by \cite{sze08}. Their calculation of Hill stability is positive and constantly increasing in time for a million years, which is consistent with our result. However, the plots are limited to a million years in the calculations of \cite{sze08} which makes the periodicity for long term oscillations for the planets to remain unclear. For the nominal case, 7 spikes in the Hill stability are noted, which indicates that the planet shows a quasi-periodic motion every 1.4 million years. When $e_{bin}$ is increased to 0.6 the period of HS function decreases and occurs every 2.5 million years. This periodicity is caused by the gravitational perturbation on the planet due to the secondary star near the pericentre of the binary. As the $e_{bin}$ is increased, the effects of secondary star on the planet is minimized because the time of interaction has been significantly reduced due to Kepler's 2nd Law, thus also reducing the period of HS function.

The HS criterion has shown that the orbital stability of the planet decreases as we increase the $e_{bin}$ of $\gamma$ Cephei. When the $e_{bin}$ was set to 0.65, the planet starts to display orbital instability at $\sim 2.25 \times 10^5$ years, a point where the HS function starts a negative trend in the time series (a cut off point shown by a vertical arrow, see Fig. \ref{fig:gCephei_HS1}c). The planet was ejected from the system at $\sim 6 \times 10^5$ years. The results supplement our previous instability prediction made from MLE analysis.

We also found that the measure of Hill stability for $\gamma$ Cephei Ab stays positive for all of our choices of $i_{pl}$.  The periodicity however decreases with increase in $i_{pl}$ values (Figs. \ref{fig:gCephei_HS2}a, \ref{fig:gCephei_HS2}b, and \ref{fig:gCephei_HS2}c). The increment in orbital inclination of the planet minimizes the effect due to the secondary star's close approach to the planet.

Similarly, the Hill stability time series stays positive for the planet in HD 196885 binary system  for the cases when $e_{bin}$ is set at 0.42 (observational value) and 0.46 (Figs. \ref{fig:HD196885_HS1}a,b). The planet displays a quasi periodic motion every 1.1 million years for our first choice of $e_{bin}$. But we found insignificant change in the periodicity for the second choice of $e_{bin}$. The periodicity was not affected because of the minute difference in $e_{bin}$, (0.04).

When the $e_{bin}$ of HD 196885 was set to 0.6, within the presumed unstable region, the HS time series demonstrates that the stability was lost within $5 \times 10^3$ years (Fig. \ref{fig:HD196885_HS1}c). For this system the ejection of the planet also occurs at $\sim 5 \times 10^3$ years of simulation time. The ejection occurs in such a short period that the HS function was not efficient enough to predict the instability.

The stability of HD 196885 Ab was not affected by our choice of orbital inclination. The system demonstrates stable orbits for $i_{pl}$ values as high as 25$^\circ$. The HS time series remains positive reflecting the established stability criterion from \citeauthor{sze08}. Just like the planet in $\gamma$ Cephei, the period of HS function for HD 196885 Ab decreases with the increase in $i_{pl}$. It would be interesting to exploit the nature of HS at higher $i_{pl}$ values when the system hits the kozai resonances, but we have limited our choice of $i_{pl}$ at 25$^\circ$ in this paper.

\subsection{Analysis of MEGNO maps} \label{sec:MEGNO_An}

\begin{figure*}
\centering
\subfloat[$e_{bin}$ = 0.4085]{\includegraphics[width=.33\linewidth]{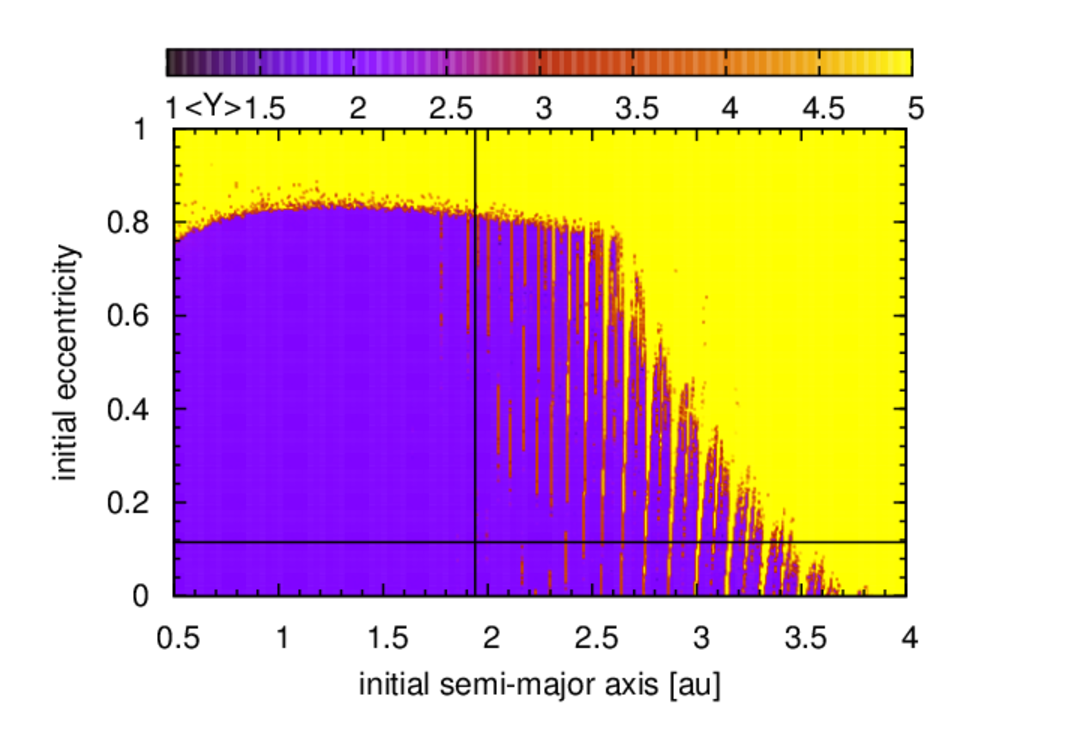}}
\subfloat[$e_{bin}$ = 0.2]{\includegraphics[width=.33\linewidth]{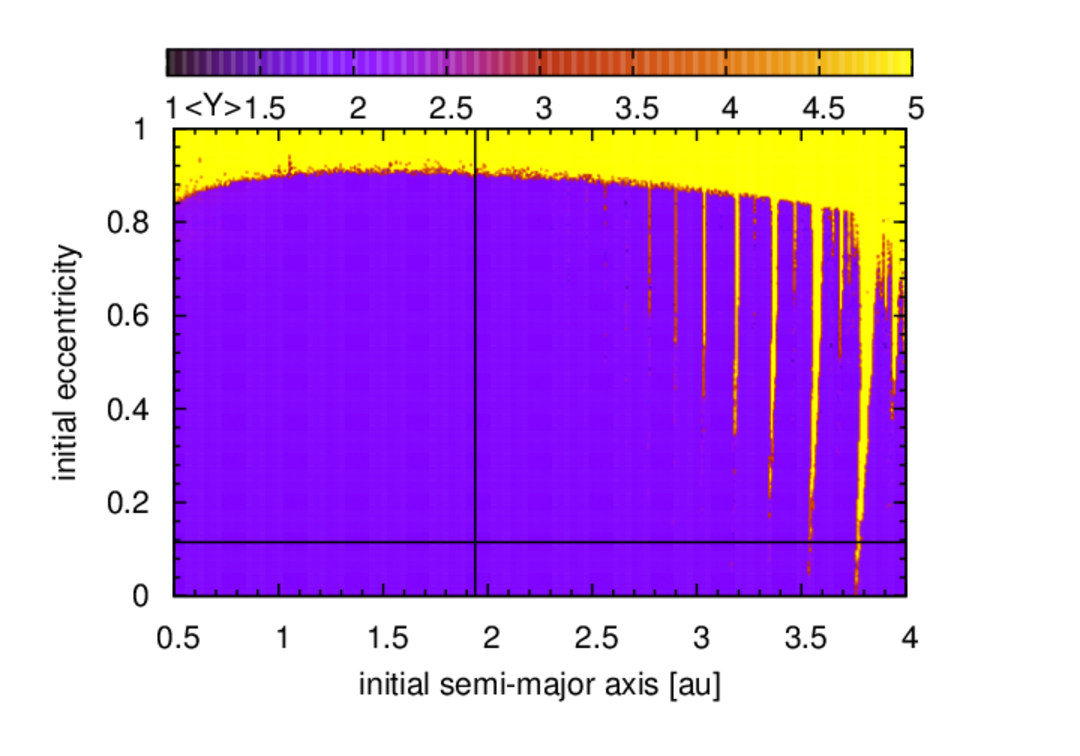}}
\subfloat[$e_{bin}$ = 0.6]{\includegraphics[width=.33\linewidth]{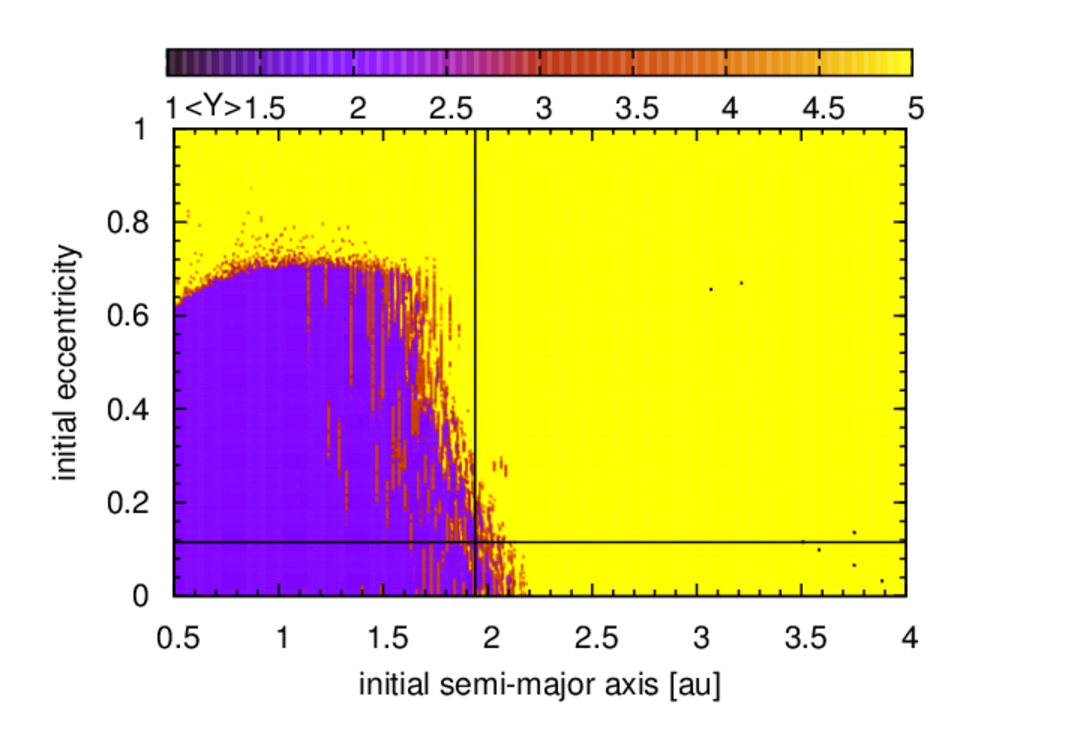}}
\caption{MEGNO Maps for the planet in $\gamma$ Cephei at $i_{pl}$ = 0$^\circ$ and various binary eccentricities simulated for $1 \times 10^5$ years.}
\label{fig:gCephei_MEGNO1}
\end{figure*}

\begin{figure*}
\centering
\subfloat[$e_{bin}$ = 0.42]{\includegraphics[width=.33\linewidth]{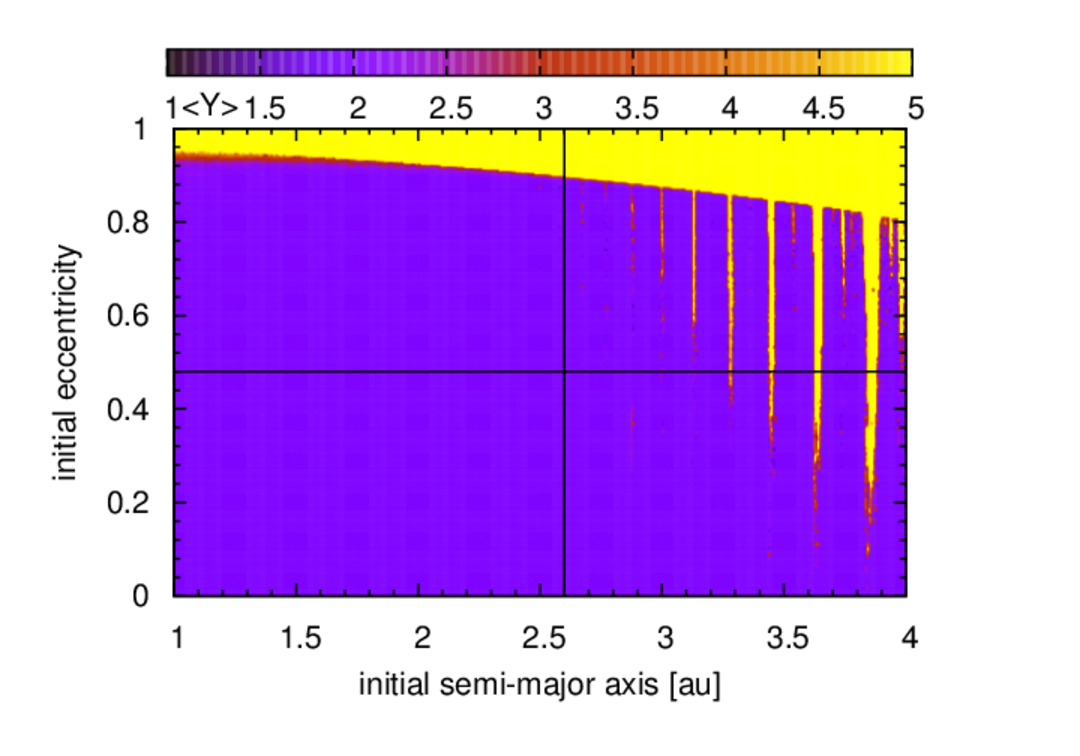}}
\subfloat[$e_{bin}$ = 0.2]{\includegraphics[width=.33\linewidth]{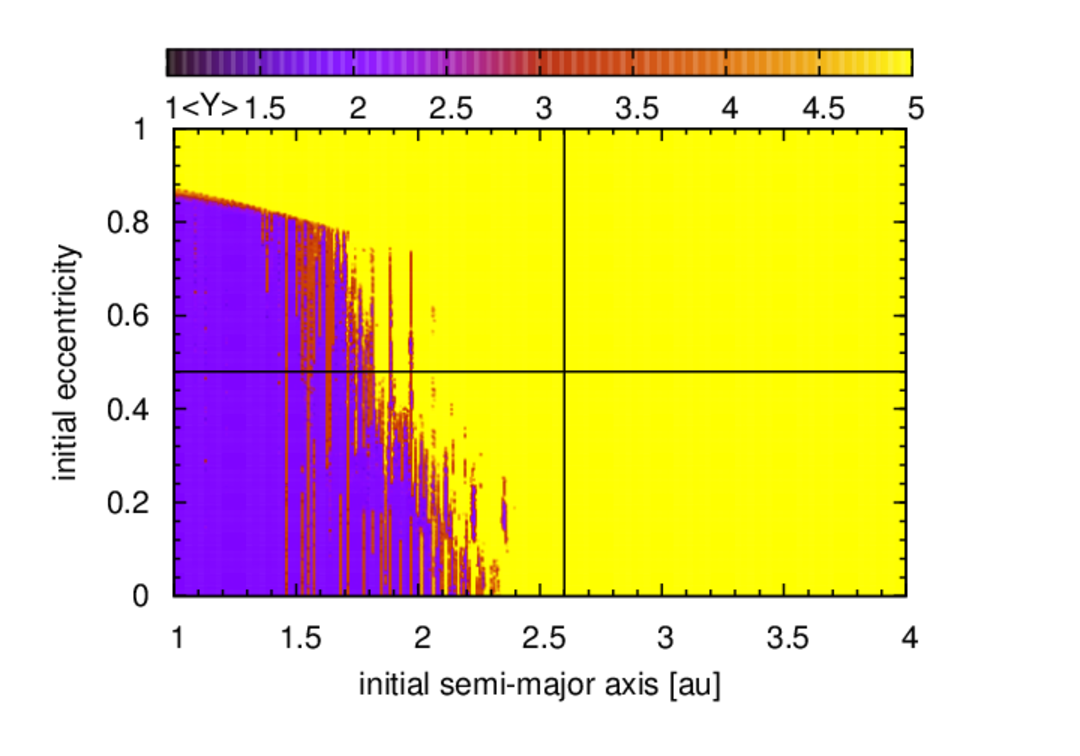}}
\subfloat[$e_{bin}$ = 0.6]{\includegraphics[width=.33\linewidth]{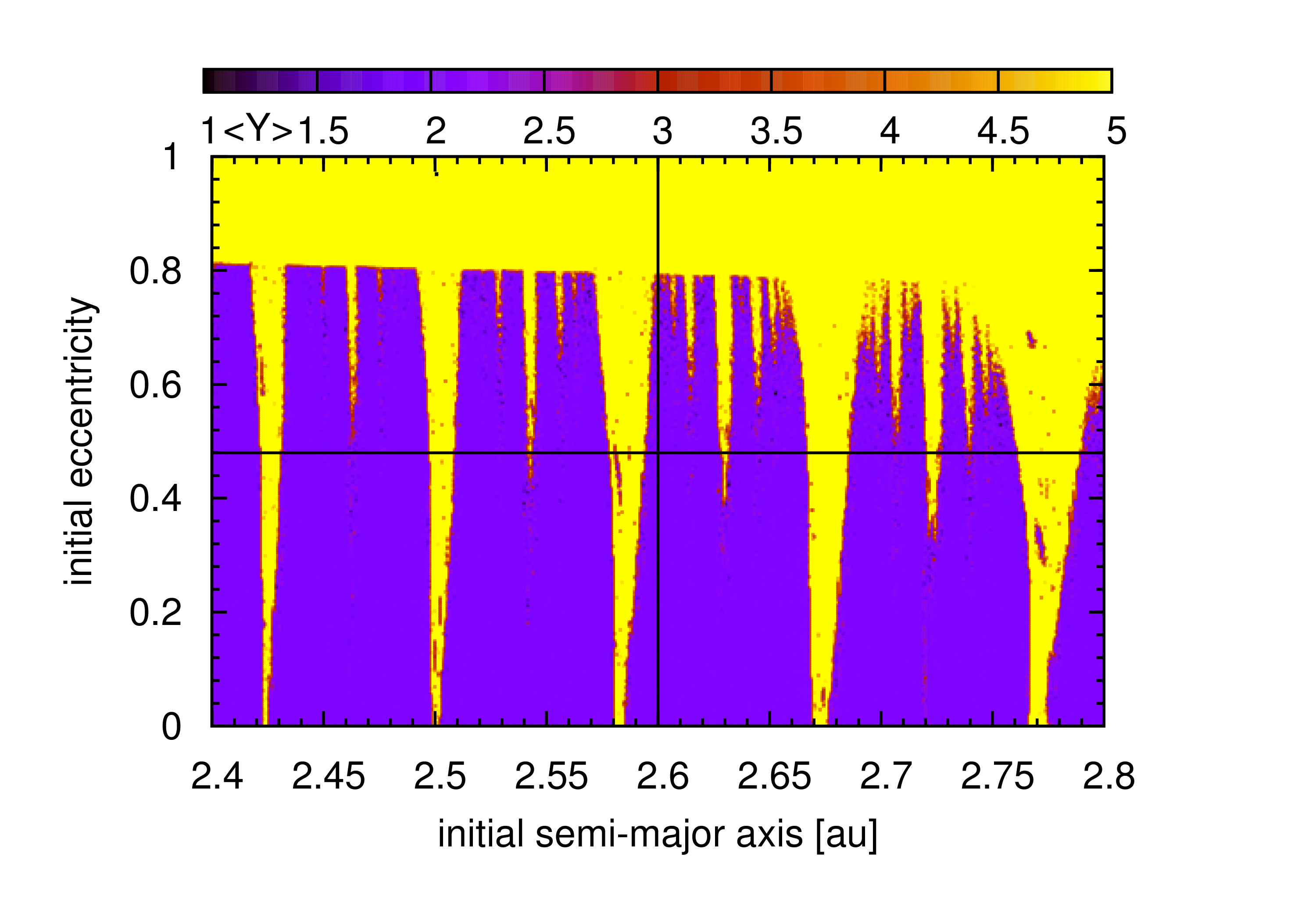}}
\caption{MEGNO Maps for the planet in HD 196885 ($i_{pl}$ = 0$^\circ$) and various binary eccentricities simulated for $1 \times 10^5$ years.}
\label{fig:HD196885_MEGNO1}
\end{figure*}

The dynamical maps of MEGNO are generated using a resolution of (500 x 300) producing 150,000 initial conditions in eccentricities and semimajor axes for the respective planets within the selected binaries. In Figs. \ref{fig:gCephei_MEGNO1} and \ref{fig:HD196885_MEGNO1} MEGNO maps for various binary eccentricities were simulated for $1 \times 10^5$ years. The cross hair in each subplot represents the osculating orbit of planet Ab for the respective binary (see Tables \ref{tab:gCephei} and \ref{tab:HD196885}). The colour bar on top of each map indicates the strength in the value of MEGNO($<Y>$). The blue colour denotes regions of quasi-periodicity and the yellow indicates regions of chaos.

For different eccentricity values of the binary in $\gamma$ Cephei, Fig. \ref{fig:gCephei_MEGNO1}, the MEGNO indicator shows a clear distinction between quasi-periodic and chaotic regions. Within the observational value, $e_{bin}$ $\sim$ 0.4085 (Fig. \ref{fig:gCephei_MEGNO1}a), the planet unmistakably demonstrates stable orbits. When the binary eccentricity, $e_{bin}$ = 0.2 (Fig. \ref{fig:gCephei_MEGNO1}b), is decreased the cross hair is completely inside the quasi-periodic region, hence increasing the orbital stability. Conversely, as the eccentricity of the binary orbit is increased, $e_{bin}$=0.6, the location of the chaotic mean-motion resonances (yellow spikes at constant semimajor axis) are shifted to lower semimajor axes of the planet (Fig. \ref{fig:gCephei_MEGNO1}c), hence decreasing the orbital stability. Studies done by \cite{hag06} for $e_{bin}$ = 0.20-0.45 at interval of 0.05 and $i_{pl}$ = 0$^\circ$ also show that the planet in $\gamma$ Cephei demonstrates stable orbits.

\begin{figure*}
\centering
\subfloat[$\gamma$ Cephei]{\includegraphics[width=.4\linewidth]{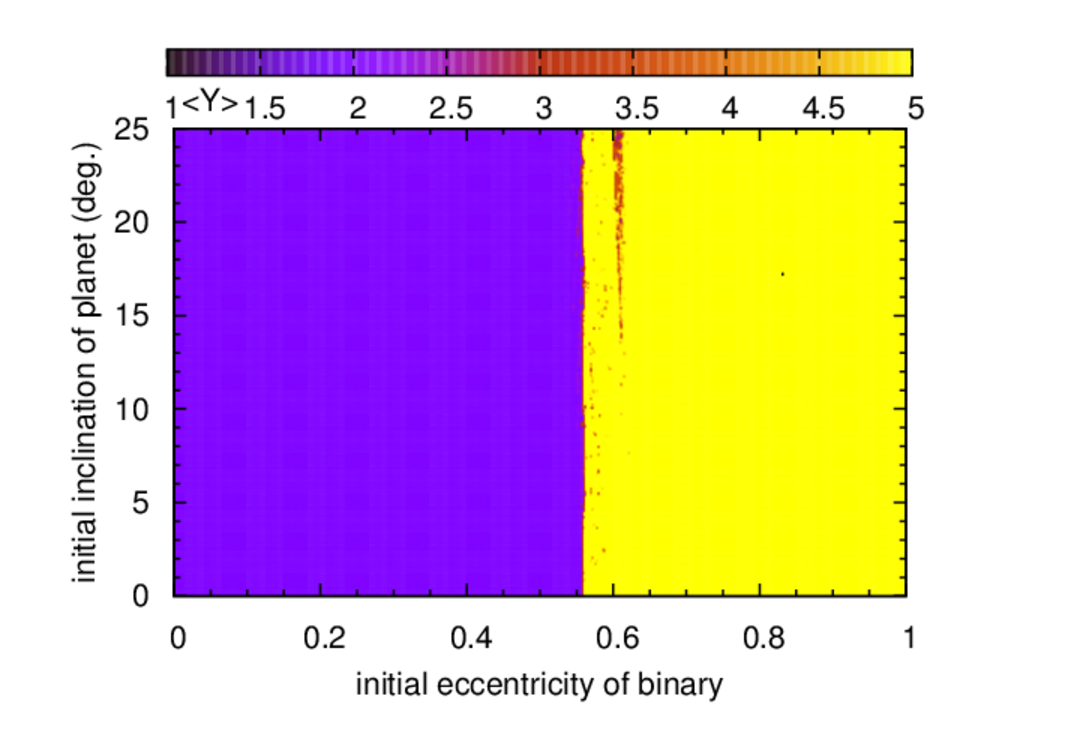}}
\subfloat[HD 196885]{\includegraphics[width=.4\linewidth]{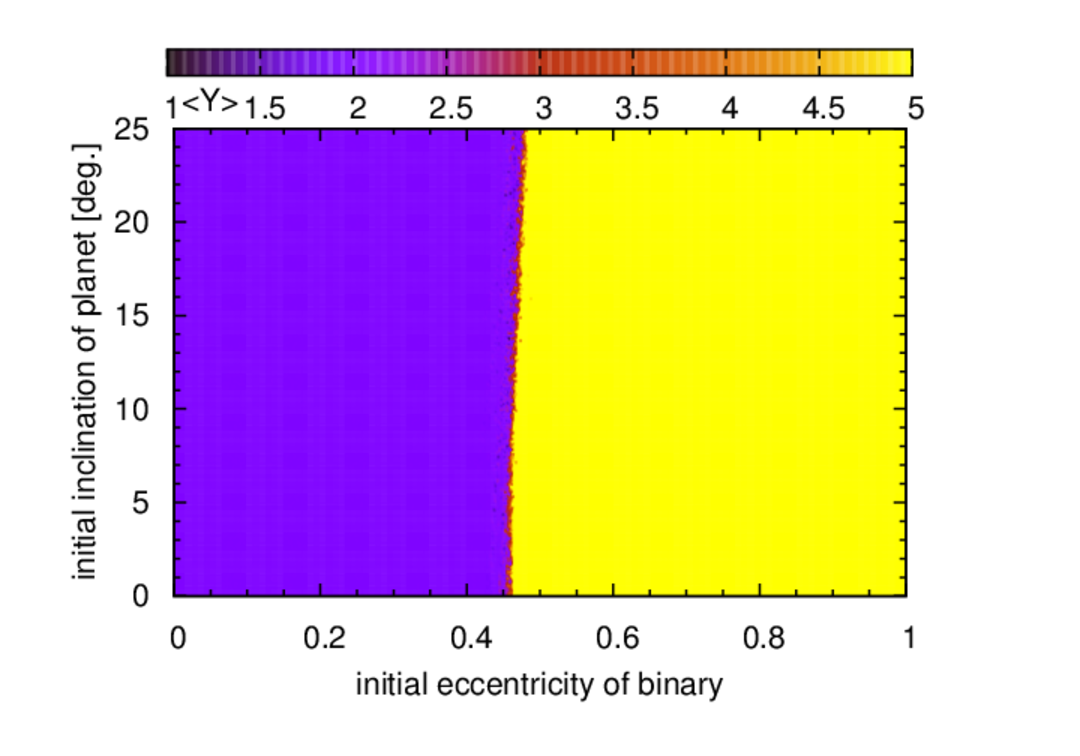}}
\caption{MEGNO Maps for the planet in (a) $\gamma$ Cephei showing variation in the ($e_{bin}$,$i_{pl}$) plane for the nominal ($a_{pl}$,$e_{pl}$) values and (b) HD 196885 showing variation in the ($e_{bin}$,$i_{pl}$) plane for its respective ($a_{pl}$,$e_{pl}$) values simulated for $1 \times 10^5$ years.}
\label{fig:MEGNO_ieb}
\end{figure*}

We also tested the stability of the planet in $\gamma$ Cephei for the given values of orbital elements by generating a global map of $e_{bin}$ vs. $i_{pl}$, where $e_{bin}$ ranges from 0.0 to 1.0 and $i_{pl}$ ranges from 0$^\circ$ to 25$^\circ$ (Fig. \ref{fig:MEGNO_ieb}). The dynamics of the planet was found unchanged for the $e_{bin}$ as high as 0.6 and any choice of $i_{pl}$ $\le$ 25$^\circ$. The system seems to show chaotic behaviour at $e_{bin}$ = 0.6 and low $i_{pl}$ values but our earlier investigation (Fig. \ref{fig:gCephei_MEGNO1}c) indicates the cross hair at the semi-chaotic region. We investigated a finer resolution window concerning this region and determined that the initial condition existed on the border of a chaotic region. Nonetheless, this confirms that for our choice of orbital inclination of the planet with the binary orbit, the planet is stable but may display chaos within the long-term ($>$ 10 Myr). Moreover our results are consistent with those obtained by \cite{hag06} where he showed the stable configuration of the system for all values of the planet's orbital inclination less than 40$^\circ$.

Figure \ref{fig:HD196885_MEGNO1}a shows a locally stable region for the planet in HD 196885 system. The cross hair in the map is located right at the edge of the stable region. A small change in semimajor axis of the planet can divert the planet towards the chaotic region loosing global stability. Similarly the location of cross hair indicates that the quasi-periodicity increases with a decrease in the $e_{bin}$, Fig. \ref{fig:HD196885_MEGNO1}b, while the system becomes chaotic with increment in $e_{bin}$, Fig. \ref{fig:HD196885_MEGNO1}c.

Figure \ref{fig:MEGNO_ieb}b, displays the case with the previously observed values of $e_{bin}$ (Fig. \ref{fig:HD196885_MEGNO1}) in a global map of $e_{bin}$ vs. $i_{pl}$ for the given values of orbital parameters of the system. It is found that the dynamics of HD 196885 Ab is unaffected by the choices of $e_{bin}$ $\le$ 0.46 and $i_{pl}$ $\le$ 25$^\circ$. \cite{giu12} showed that the system is mostly stable when the planetary orbit is nearly coplanar or highly inclined orbits near the Lidov-Kozai equilibrium point, ($i_{pl}$ = 47$^\circ$). We did not test the system for $i_{pl}$$>$25$^\circ$, however, our result is consistent to the previous work with respect to MLE and HS time series of the respective inclination range.

\begin{figure*}
\centering
\subfloat{\includegraphics[width=.3\linewidth]{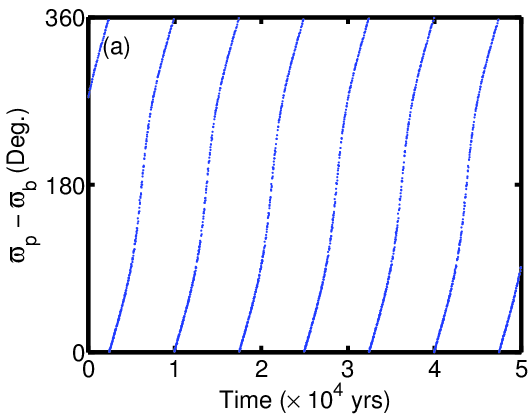}}
\subfloat{\includegraphics[width=.3\linewidth]{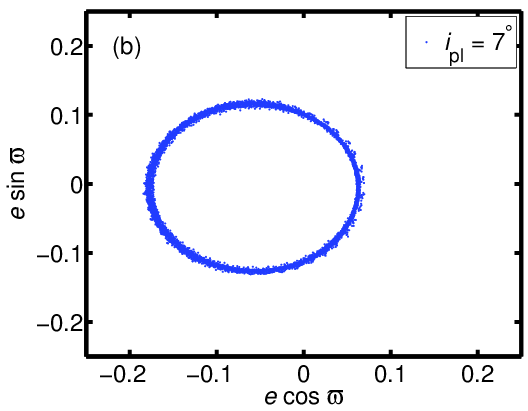}}
\subfloat{\includegraphics[width=.3\linewidth]{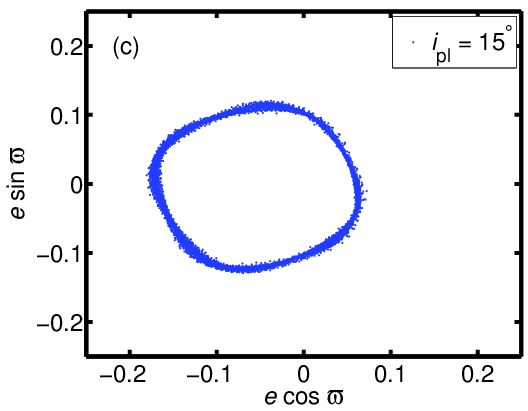}}
\\*
\subfloat{\includegraphics[width=.3\linewidth]{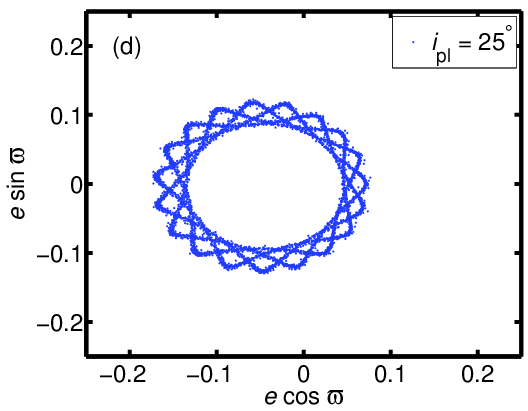}}
\subfloat{\includegraphics[width=.3\linewidth]{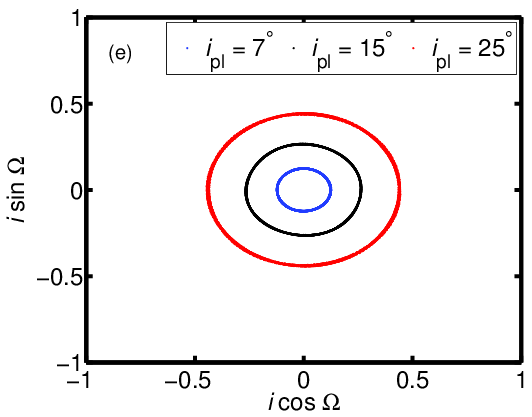}}
\caption{(a) Difference between the osculating longitude of pericentres ($\varpi$) for $\gamma$ Cephei for $5 \times 10^4$ years. (b), (c), and (d) Evolution of osculating values ($e \cos \varpi$, $e \sin \varpi$) for different orbital inclinations. (e) Evolution of osculating values ($i$ cos $\Omega$, $i \sin \Omega$) for different orbital inclination.}
\label{fig:gCephei_w}
\end{figure*}

\begin{figure*}
\centering
\subfloat{\includegraphics[width=.3\linewidth]{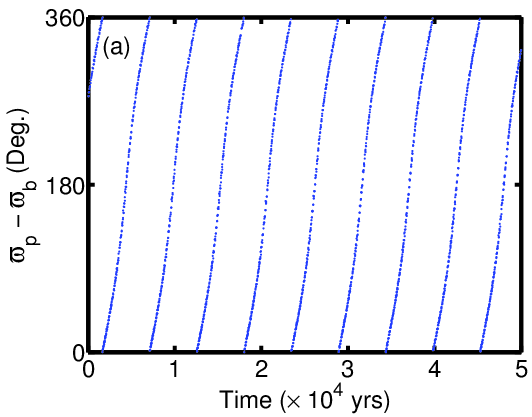}}
\subfloat{\includegraphics[width=.3\linewidth]{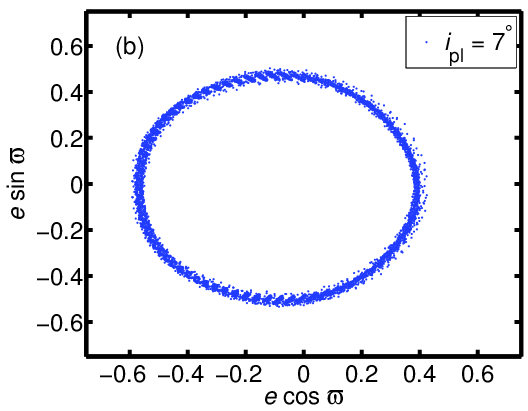}}
\subfloat{\includegraphics[width=.3\linewidth]{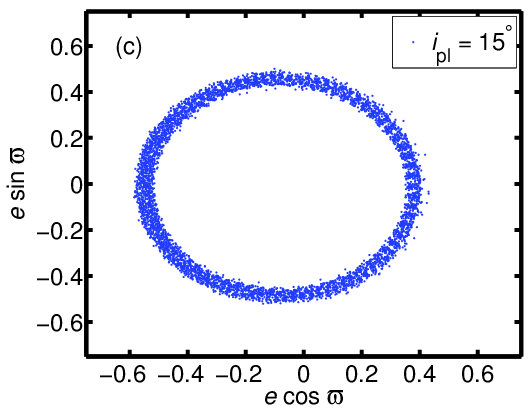}}
\\*
\subfloat{\includegraphics[width=.3\linewidth]{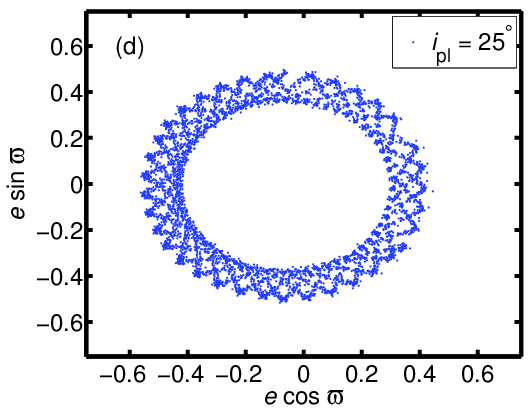}}
\subfloat{\includegraphics[width=.3\linewidth]{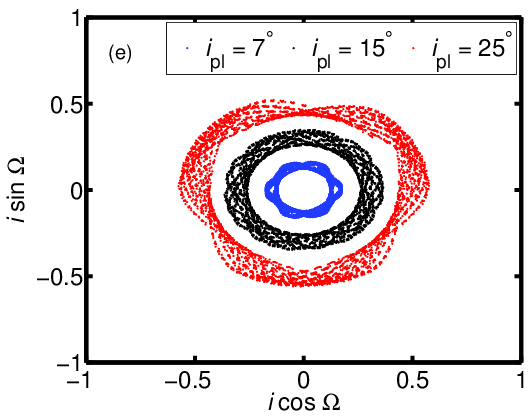}}
\caption{(a) Difference between the osculating longitude of pericentres ($\varpi$) for HD 196885 for $5 \times 10^4$ years. (b), (c), and (d) Evolution of osculating values ($e \cos \varpi$, $e \sin \varpi$) for different orbital inclinations. (e) Evolution of osculating values ($i \cos \Omega$, $i \sin \Omega$) for different orbital inclination.}
\label{fig:HD196885_w}
\end{figure*}

\subsection{Evolution of Osculating parameters}

Considering the orbital parameters of $\gamma$ Cephei and HD 196885 from the observations, the evolution of argument of periapsis ($\omega$) and longitude of ascending node ($\Omega$) are studied. In Figs. \ref{fig:gCephei_w}a and \ref{fig:HD196885_w}a we have shown the difference between longitude of periapsis between the planet and the binary (where $\varpi$ = $\omega$ + $\Omega$). The longitude of periapsis oscillates and is bound within 0$^\circ$ to 360$^\circ$ which exhibits an apocentric libration phenomenon \citep{mur99}.

Plots of the osculating values, $e \cos \varpi$ and $e \sin \varpi$, for the planet's nominal choice of parameters indicated no change in the osculating parameters of both systems. When the orbital inclination was increased (up to 25$^\circ$) the precession rates were noticeably increased (see Figs. \ref{fig:gCephei_w}b, c, d and \ref{fig:HD196885_w}b, c, d), an effect similar to period-doubling bifurcation \citep{ott93} leading to chaos. This change is not noticed directly in LE time series for $\gamma$ Cephei (Fig. \ref{fig:gCephei_LE}d) but we do observe variation (different converging rate) in the LE  time series for HD 196885 (Fig. \ref{fig:HD196885_LE}d) and variation (decreasing periodicity) in HS time series for both of the systems (Figs. \ref{fig:gCephei_HS2} and \ref{fig:HD196885_HS2}).

Figures \ref{fig:gCephei_w}e and \ref{fig:HD196885_w}e display plots of the osculating parameters,  $i \cos \Omega$ and $i \sin \Omega$, for different cases of orbital inclination. For the coplanar case, we would have a point-circle at the origin and the circle would grow larger for larger values of $i_{pl}$, as seen in these diagrams. The osculating parameters with respect to inclination indicates regular motion in the case of $\gamma$ Cephei. But in the case of HD 196885 significant variations occur with increases in inclination (Fig. \ref{fig:HD196885_w}e).

\subsection{Reliability of Hill stability function}

Following the work done by \cite{sze08}, we used the Hill stability criterion to study the orbital stability of planets in stellar binaries ($\gamma$ Cephei and HD 196885). We also used the previously well known chaos indicators, MLE and the MEGNO maps to study the same systems. The maximum Lyapunov exponent (Figs. \ref{fig:gCephei_LE}a and \ref{fig:HD196885_LE}a) and the cross hair in the quasi periodic region of MEGNO map (Figs. \ref{fig:gCephei_MEGNO1}a and \ref{fig:HD196885_MEGNO1}a) demonstrate a stable configuration for given parameters in each of the considered systems. The positive value and non-decreasing global trend of Hill stability time series (Figs. \ref{fig:gCephei_HS1}a and \ref{fig:HD196885_HS1}a) also provides the necessary evidence of stable system. The HS time series indicates that the systems are quasi-periodic and the periodicity decreases as the planet's inclination and binary eccentricity is increased. Since the calculation for MLE is limited to 1 million years, we are unable to notice any periodicity in the time series as seen in the HS time series but it does also indicate trends toward stability. Similarly, MEGNO maps do not demonstrate any changes in stability when the respective planet's inclination increased up to 25$^\circ$, nonetheless, it shows that the planets in both of the system exist within a stable configuration.

The stability of the planets is tested for highly eccentric binary orbits using MLE and MEGNO maps (Figs. \ref{fig:gCephei_LE}c and \ref{fig:HD196885_LE}c and \ref{fig:MEGNO_ieb}). These figures clearly demonstrate unstable orbits for chosen $e_{bin}$ values. Similar tests done by using HS time series (Figs. \ref{fig:gCephei_HS1}c and \ref{fig:HD196885_HS1}c) produce similar results regarding stability of the system.

The HS time series supplements two of our results from MLE and MEGNO for the cases when orbits are periodic and chaotic, thus establishing itself as a reliable stability criterion. For the planet in $\gamma$ Cephei (see Figs. \ref{fig:gCephei_LE}c and \ref{fig:gCephei_HS1}c) the HS function indicated an instability within the planetary orbit on an equal simulation timescale compared to MLE (see section \ref{sec:MLE_An}). However it took a slightly longer simulation time to predict instability as compared to MLE for the case of HD 196885. For the system where the planet ejects in relatively short time ($\sim 5 \times 10^3$ years in this case) MLE is more efficient to predict regular or chaotic dynamical systems and the HS function seems to be an inefficient indicator (Figs. \ref{fig:HD196885_LE}c and \ref{fig:HD196885_HS1}c).

There are limitations to the definition of HS as well.  Most notably HS is defined for S-Type orbits only. Considering this limitation, it may be best suited in future work to the study of hypothetical or observed moons around gas giants as observed S-Type planet populations remain modest \citep{kip12}.  However, the populations of multi-planet single star systems are rapidly increasing with the use of Transit Timing Variations (see \cite{lis11,lis13} for recent results) and the HS criterion would provide adequate stability determinations for similar systems.

\section{Conclusions}
We have applied various chaos indicator techniques in order to study the dynamics of S-type extrasolar planets in the binaries that are less than 25 AU apart. With the time series obtained from the maximum Lyapunov exponent, the Hill stability function, and the maps from the MEGNO indicator, we have shown that both the systems exist within a stable configuration for given parameters. Using these chaos indicators, we have also tested the orbital stability of the system for various choices of binary eccentricity and planet's orbital inclination.

The calculated MLE and HS time series for the planets in both systems for different values of planet's orbital inclination, $i_{pl}$ = 0$^\circ$, 7$^\circ$, 15$^\circ$, 25$^\circ$, and given $e_{bin}$ value display the orbital stability of the system. For the given values of orbital parameters it is found that the planet in $\gamma$ Cephei/HD 196885 can maintain stability for $e_{bin}$ as high as 0.6/0.46. Similar studies made by using the global MEGNO maps of $e_{bin}$ vs. $i_{pl}$ for given orbital parameters concur with our results from MLE and HS and demonstrate that the planet in $\gamma$ Cephei/HD 196885 can maintain stability for $i_{pl}\le$ 25$^\circ$ and $e_{bin}\le$ 0.6/0.46.

The MEGNO chaos indicator has been effective in determining the quasi-periodic regions. The location of eccentricity-semimajor axis cross hairs in the MEGNO maps for the planets in  $\gamma$ Cephei and HD 196885 systems (Figs. \ref{fig:gCephei_MEGNO1}a and \ref{fig:HD196885_MEGNO1}a) are located well inside the quasi periodic region (blue) and in the teeth between the chaotic and quasi periodic regions, respectively. This resembles the global and local stability of the planets. These planets do not survive if the binary orbits are highly eccentric (see Figs. \ref{fig:MEGNO_ieb}a,b).

The Hill stability time series for a planet is successfully measured using the numerical integration of orbital parameters and conversion into a potential related to the elliptic restricted three body problem. The measure of the Hill stability of the planets in $\gamma$ Cephei and HD 196885 has shown changes in periodicity with the variation of $i_{pl}$ and $e_{bin}$ values. These periodicities are believed to have originated due to the differences in the time of interaction on the planet due to the secondary star near time of periastron passage for the binary.

A concise study of evolution of osculating parameters show that they have insignificant variation in both of the systems for the nominal case. But, as we increase the $i_{pl}$ values up to 25$^\circ$ the osculating values ($e \cos \varpi$ and $e \sin \varpi$) evolve causing the precession rate in $\gamma$ Cephei and HD 196885 to increase. In the set of osculating parameters ($i \cos \Omega$ and $i \sin \Omega$) the inclination varies with a very small amplitude while the longitude of ascending node rotates in the full range [$0^{\circ},360^{\circ}$]. In case of HD 196885 these parameters display significant evolution which provides possible evidence towards chaotic behaviour in this regime.

Aside from the dynamical analysis of planets in stellar binaries, we are able to successfully test the reliability of Hill stability against the results obtained from both MLE and MEGNO. Direct comparison of stability shows that the Hill stability test can be set as one of the three stringent criteria in the study of stable/unstable nature of a planetary orbit. Our results show that the HS indicator is comparable, in the context of determining the orbital stability, to other well known indicators. Like MLE and MEGNO, it has consistently predicted the stable/unstable nature of planets in binaries. Calculations show that the HS function predicts instability of orbits in comparable time with LE (case of $\gamma$ Cephei). Also, unlike MEGNO maps, HS time series cannot produce definite quasi-periodic, chaotic or decreasing stability regions without considering a map of a larger parameter space.

\emph{\textbf{Acknowledgement}}: SS and BQ would like to thank department of Physics UT Arlington, Zdzislaw Musielak and Manfred Cuntz for their continuous support and guidance. TCH gratefully acknowledges financial support from the Korea Research Council for Fundamental Science and Technology (KRCF) through the Young Research Scientist Fellowship Program and financial support from KASI (Korea Astronomy and Space Science Institute) grant number 2013-9-400-00. Numerical computations were partly carried out using the SFI/HEA Irish Centre for High-End Computing (ICHEC) and the PLUTO computing cluster at the Korea Astronomy and Space Science Institute. Astronomical research at the Armagh Observatory is funded by the Northern Ireland Department of Culture, Arts and Leisure (DCAL). We would like to thank Cezary Migaszewski for his useful comments and suggestions which have brought a lot of improvements to this work.

\bibliographystyle{mn2e}
\bibliography{references}

\end{document}